\documentclass[aps,pre,twocolumn,showpacs]{revtex4}
\usepackage{epsfig}
\usepackage{times}
\begin{document}

\title{Social diversity and promotion of cooperation in the spatial prisoner's dilemma game}

\author{Matja{\v z} Perc$^1$ and Attila Szolnoki$^2$}
\affiliation
{$^1$Department of Physics, Faculty of Natural Sciences and Mathematics, University of \\ Maribor, Koro{\v s}ka cesta 160, SI-2000 Maribor, Slovenia \\
$^2$Research Institute for Technical Physics and Materials Science, 
P.O. Box 49, H-1525 Budapest, Hungary}

\begin{abstract}
The diversity in wealth and social status is present not only among humans, but throughout the animal world. We account for this observation by generating random variables that determine the social diversity of players engaging in the prisoner's dilemma game. Here the term social diversity is used to address extrinsic factors that determine the mapping of game payoffs to individual fitness. These factors may increase or decrease the fitness of a player depending on its location on the spatial grid. We consider different distributions of extrinsic factors that determine the social diversity of players, and find that the power-law distribution enables the best promotion of cooperation. The facilitation of the cooperative strategy relies mostly on the inhomogeneous social state of players, resulting in the formation of cooperative clusters which are ruled by socially high-ranking players that are able to prevail against the defectors even when there is a large temptation to defect. To confirm this, we also study the impact of spatially correlated social diversity and find that cooperation deteriorates as the spatial correlation length increases. Our results suggest that the distribution of wealth and social status might have played a crucial role by the evolution of cooperation amongst egoistic individuals. 
\end{abstract}


\maketitle\section{Introduction}
The prisoner's dilemma game, consisting of cooperation and defection as the two competing strategies, is considered a paradigm for studying the emergence of cooperation between selfish individuals \cite{axelrod_84}. Since the game promises a defecting individual the highest income if facing a cooperator the prevalence of cooperation within this theoretical framework presents a formidable challenge. Indeed, the classical well-mixed prisoner's dilemma game completely fails to sustain cooperation \cite{hofbauer_98}, which is often at odds with reality where mutual cooperation may also be the final outcome of the game \cite{wilkinson_n84}. A seminal theoretical mechanism for cooperation within the prisoner's dilemma game was introduced by Nowak and May \cite{nowak_n92b}, who showed that the spatial structure and nearest neighbor interactions enable cooperators to form clusters on the spatial grid and so protect themselves against exploitation by defectors. A decade later, this theoretical prediction has been confirmed by biological experiments \cite{kerr_n02}. Nonetheless, the somewhat fragile ability of the spatial structure to support cooperation \cite{huberman_pnas93}, along with the difficulties associated with payoff rankings in experimental and field work \cite{milinski_prsb97}, has made it a common starting point for further refinements of cooperation facilitating mechanisms. In particular, the specific topology of networks defining the interactions among players has recently received substantial attention \cite{abramson_pre01}, and specifically scale-free graphs \cite{barabasi_s99} have been recognized as extremely potent promoters of cooperative behavior in the prisoner's dilemma as well as the snowdrift game \cite{santos_prl05}. Although the promotion of cooperation by the scale-free topology has been found robust on several factors \cite{poncela_njp07}, the mechanism has recently been contested via the introduction of normalized payoffs or so-called participation costs \cite{wu_pa07}. Moreover, the interplay between the evolution of cooperation as well as that of the interaction network has also been studied \cite{zimmermann_pre05}, and it has been discovered that intentional rewiring in accordance with the preference and fitness of each individual as well as simple random rewiring of the interaction network might both have a beneficial effect on the evolution of cooperation in the prisoner's dilemma and the snowdrift game. For a comprehensive review of this field of research see \cite{szabo_pr07}.

Besides studies addressing network complexity as a somewhat direct extension of \cite{nowak_n92b}, several approaches have also been proposed that warrant the promotion of cooperation within nearest neighbor interactions. Examples include strategic complexity \cite{hauert_s02}, direct and indirect reciprocity \cite{nowak_n98}, asymmetry of learning and teaching activities \cite{szolnoki_epl07}, random diffusion of agents on the grid \cite{vainstein_pre01}, as well as fine-tuning of noise and uncertainties by strategy adoption \cite{perc_pre07a}. The impact of asymmetric influence, introduced via a special player with a finite density of directed random links to others, on the dynamics of the prisoner's dilemma game on small-world networks has also been studied \cite{kim_2002}. It is worth noting that stochasticity in general, either being introduced directly or emerging spontaneously due to finite population sizes \cite{traulsen_prl05}, has been found crucial by several aspects of various evolutionary processes.

In this paper, we wish to extend the scope of stochastic effects on the evolution of cooperation in the spatial prisoner's dilemma game by introducing the social diversity of players as their extrinsically determined property. This is realized by introducing scaling factors that determine the mapping from game payoffs to the fitness of each individual. The scaling factors represent extrinsic differences amongst players, and as such determine their social status and the overall social diversity on the spatial grid. Importantly, the scaling factors are drawn randomly from different distributions and are determined only once at the beginning of the game. Positive scaling factors increase the magnitude of payoffs a particular player is able to exchange, while negative factors have the opposite effect. It is important to note that the average of all scaling factors is exactly zero, so that there is no net contribution of social diversity to the total payoff of the population. Presently, we consider scaling factors drawn from the uniform, exponential, and scale-free distribution, and thus distinguish three different cases of social diversity. We investigate how the introduction of social diversity, and in particular its distribution and amplitude, affect the evolution of cooperation amongst players on the spatial grid. We report below that the cooperation is enhanced markedly as the amplitude of social diversity increases, and moreover, that the biggest enhancement is obtained if the social diversity follows a power-law distribution. We attribute the enhancement of cooperation to the emergence of cooperative clusters, which are controlled by high-ranking players that are able to prevail against the defectors even if the temptation to defect is large. Indeed, the role of these clusters is similar to the role of pure cooperator neighborhoods around hubs on scale-free networks \cite{gomez-gardenes_prl07}, or to the part of imitating followers surrounding master players in the enhanced teaching activity model \cite{szolnoki_epl07}, as will be clarified later. To highlight the importance of the strongly diverse social rank, we also study the impact of spatially correlated social diversity, and find that the facilitation of the cooperative strategy deteriorates as the correlation length increases. Thus, a finite spatial correlation of social diversity hinders the formation of strong cooperative clusters, in turn lending support to the validity of the reported cooperation-facilitating mechanism in the spatial prisoner's dilemma game.

The remainder of this paper is structured as follows. Section II is devoted to the description of the spatial prisoner's dilemma game and the properties of scaling factors determining social diversity of players on the spatial grid. In Section III we present the results of numerical simulations, and in Section IV we summarize the results and outline some biological implications of our findings.

\section{Model definition}
We consider an evolutionary two-strategy prisoner's dilemma game with players located on vertices of a two-dimensional square lattice of size $L \times L$ with periodic boundary conditions. Each individual is allowed to interact only with its four nearest neighbors, and self-interactions are excluded. A player located on the lattice site $i$ can change its strategy $s_i$ after each full iteration cycle of the game. The performance of player $i$ is compared with that of a randomly chosen neighbor $j$ and the probability that its strategy changes to $s_j$ is given by \cite{szabo_98}
\begin{equation}
W ( s_i \leftarrow s_j) = \frac{1}{1+ \exp[(P_i - P_j) / K]} \,\,,
\label{eq:prob}
\end{equation} 
where $K=0.1$ characterizes the uncertainty related to the strategy adoption process, serving to avoid trapped conditions and enabling smooth transitions towards stationary states. The payoffs $P_i$ and $P_j$ of both players acquired during each iteration cycle are calculated in accordance with the standard prisoner's dilemma scheme \cite{nowak_n92b}, according to which the temptation to defect $T = b > 1$, reward for mutual cooperation $R = 1$, punishment for mutual defection $P = 0$, and the sucker's payoff $S = 0$. Importantly, throughout this work the original prisoner's dilemma payoff ranking ($T > R > P > S$) is applicable since the introduction of social diversity via scaling factors only acts as a mapping of the original payoffs to individual fitness, as will be described next.

\begin{figure}
\centering
\includegraphics[width=8.5cm]{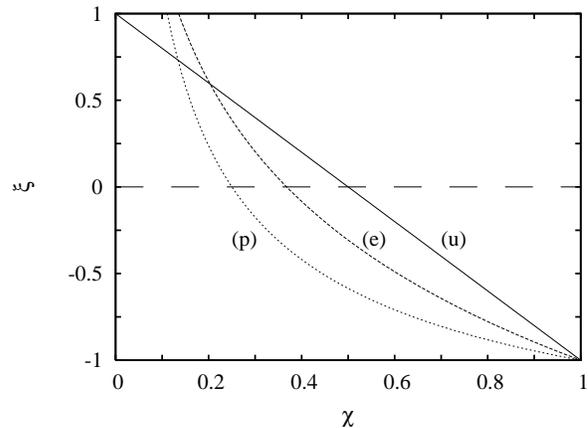}
\caption{Distributions of scaling factors ($\xi$) originating from the uniformly distributed random variable ($\chi$). The three functions correspond to uniform (u), exponential (e), and power-law (p) distribution when $a = 1$. For easier comparison only the $[-1, 1]$ interval of $\xi$ is plotted.}
\label{fig:fig1}
\end{figure}

\begin{figure}
\centering
\includegraphics[width=8.5cm]{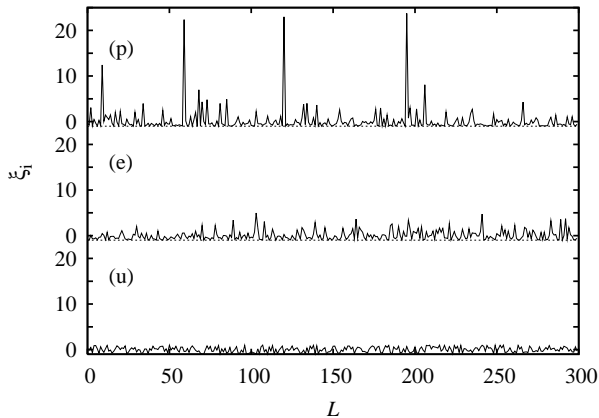}
\caption{Spatial distributions of social diversity obtained for $a = 1$ when uniform (u), exponential (e), and power-law (p) distributed values of scaling factors were used. Lines show an exemplary cross-section of the two-dimensional grid.}
\label{fig:fig2}
\end{figure}

To introduce social diversity we use rescaled payoffs of the form $X' = X(1+\xi)$, where $X$ is either $T$, $R$, $P$ or $S$, $\xi = \min(\xi_i, \xi_j)$, and $\xi_i$ is a scaling factor drawn randomly from a given distribution for each participating player $i$ only once before the start of the simulation. Note that the minimum of the two random scaling factors involved at every instance of the game is used for payoffs of both involved players $i$ and $j$ to ensure that the prisoner's dilemma payoff ranking is preserved. It is worth mentioning that the maximum of both values would have had the same effect, but below presented results are virtually independent of this technicality, and moreover, it seems reasonable to assume that the higher-ranking player (the one with the larger value of $\xi$) will adjust to the weaker one since the latter simply cannot match the stakes of the game otherwise. Within this study we consider the uniform (also known as rectangular), exponential, and power-law distributed social diversity defined by the following functions:
\begin{eqnarray}
\xi &= a &(-2 \,\chi + 1) \\
\xi &= a &(-\log\chi - 1) \\
\xi &= a &(\chi^{-1/n} - {n \over {n-1}}) \,\,\,\, \mbox{where} \, 2\le n\in N. 
\label{eq:dist}
\end{eqnarray} 
Here $\chi$ are uniformly distributed random numbers from the unit interval, and $\int_0^1 \xi(\chi) d\chi = 0$ in all cases, so that the average of $\xi$ over all the players is zero. It is easy to see that the largest difference between the exponential and power-law function can be obtained if $n = 2$ in the latter, as shown in Fig.~\ref{fig:fig1}. Henceforth we use $n=2$ when the power-law distribution is applied. The parameter $a$ determines the amplitude of undulation of the scaling factors, and hence the dispersion of social diversity, and can occupy any value from the unit interval. In particular, $a = 0$ returns the original payoffs, whereas $a = 1$ is the maximally allowed value that still preserves $(1+\xi_i) \geq 0$ for all $i$, thus preventing possible violations of the prisoner's dilemma payoff ranking. Note that $(1+\xi_i) < 0$ could induce $R > T$, which would directly violate the rules of the prisoner's dilemma game. The spatial distributions of the three considered cases obtained by $a = 1$ are demonstrated in Fig.~\ref{fig:fig2} where cross-sections with linear system size $L = 300$ are plotted. Clearly, the uniform distribution provides the gentlest dispersion of social diversity, whereas the largest segregation of players is warranted by the power-law distributed scaling factors where some values are very high at the expense of extended regions of very small $\xi$ resulting in these players having much smaller payoffs as compared to the non-scaled case. The exponential distribution of scaling factors yields a dispersion of social diversity that is between the uniform and the power-law case, as can be inferred from Fig.~\ref{fig:fig2}, as well as indirectly also from Fig.~\ref{fig:fig1}.

\begin{figure}
\centering
\includegraphics[width=8.5cm]{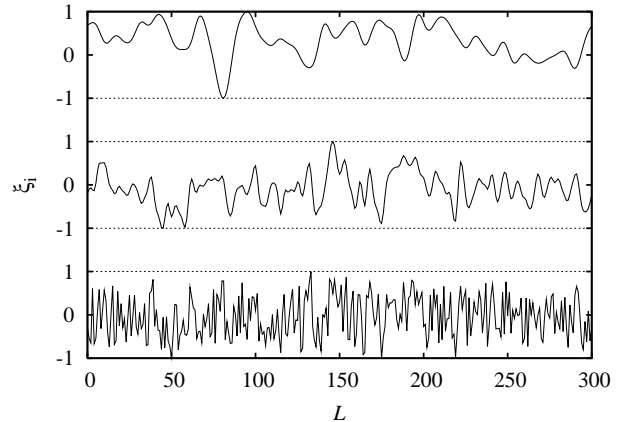}
\caption{Spatial distributions of social diversity constituted by randomly drawn scaling factors from a uniform distribution by $a = 1$ and spatial correlation length $\lambda = 1$ (bottom), $\lambda = 3$ (middle), and $\lambda = 7$ (top). Lines show an exemplary cross-section of the two-dimensional grid.}
\label{fig:fig3}
\end{figure}

In order to explain the main features of the reported results, we also use spatially correlated social diversity so that the scaling factors $\xi_i$ satisfy the correlation function $\langle \xi_i \xi_j\rangle \propto \exp(\mid i - j\mid / \lambda)$, where $\lambda$ is the spatial correlation length. In this case we constrain our study to the case where $\xi_i$ are drawn from a uniform distribution within the interval $[-1, 1]$. An efficient algorithm for the generation of spatially correlated random numbers with a prescribed correlation length is given in \cite{traulsen_04}. The effect of different values of $\lambda$ on the scaling factors determining social diversity is demonstrated in Fig.~\ref{fig:fig3}. Clearly, the random undulations become more and more correlated across neighbors on the spatial grid as $\lambda$ increases.

Before the start of each game simulation, both strategies populate the spatial grid uniformly and the scaling factors are drawn randomly from a given distribution to determine the social diversity of participating players. After these initial conditions are set, the spatial prisoner's dilemma game is iterated forward in time using a synchronous update scheme, thus letting all individuals interact pairwise with their four nearest neighbors. After every iteration cycle of the game, all players simultaneously update their strategy according to Eq.~\ref{eq:prob}. To avoid finite-size effects, the simulations were carried out for a population of $300 \times 300$ players, and the equilibrium frequencies of cooperators were obtained by averaging over $10^4$ iterations after a transient of $10^5$ iteration cycles of the game. The figures showing values of cooperator densities on the spatial grid ($F_C$) resulted from an average over $30$ simulations with different realizations of social diversity as specified by the appropriate parameters. These simulation parameters yield at least $\pm 3$ \% accurate values of $F_C$ in all figures.

\section{Numerical results}
\begin{figure*}
\centering
\includegraphics[width=14cm]{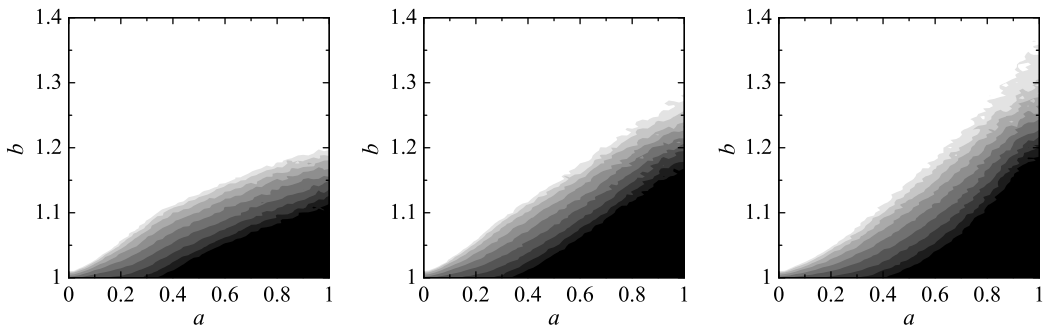}
\caption{Color-coded $F_C$  for the uniform (left), exponential (middle), and power-law (right) distributed social diversity on the $b - a$ parameter space. The color scale is linear, white depicting $0.0$ and black $1.0$ values of $F_C$.}
\label{fig:fig4}
\end{figure*}

Next, we study how different distributions of social diversity affect the evolution of cooperation via numerical simulations of the above-described spatial prisoner's dilemma game. Figure~\ref{fig:fig4} features color-coded $F_C$ for all three types of social diversity in the relevant $b - a$ parameter space. Evidently, regardless of the distribution, increasing values of $a$ clearly promote cooperation as the threshold of cooperation extinction increases from $b = 1.01$ in the absence of social diversity to $b = 1.20$ (uniformly distributed social diversity), $b = 1.28$ (exponentially distributed social diversity), and $b = 1.34$ (power-law distributed social diversity). Moreover, there always exists a broad range in the parameter space within which cooperators rule completely; a non-existent feature if $a = 0$.

We argue that the above-reported facilitation of cooperation is due to the induced inhomogeneous social state of players that fosters cooperative clusters around players with the largest values of $\xi_i$. Indeed, as soon as cooperators overtake these prime spots of the grid they start to spread due to their cluster-forming nature. Note that the latter feature is not associated with defectors who therefore fail to take the same advantage out of social diversity, and are thus defeated. A similar behavior underlies also the cooperation-facilitating mechanism reported for the scale-free networks, where the players with the largest connectivity dominate the game. Since cooperators are much better equipped for permanently sustaining the occupation of hubs, the scale-free networks provide a unifying framework for the evolution of cooperation \cite{santos_prl05}. An even better similarity can be established with the model incorporating the inhomogeneous teaching 
activity \cite{szolnoki_epl07}. In the latter, some players are blocked and therefore unable to donate their strategies, which results in homogeneous cooperative domains around players with full teaching capabilities. In the presently studied model, the introduction of social diversity partly results in a similar suppression of selected players, especially so for the power-law distributed case where the majority of players are low-ranking, as demonstrated in Fig.~\ref{fig:fig1}. These players can form an obedient domain around a high-ranking player and so prevail against defectors. On the other hand, a high-ranking defector will be weakened by the low-ranking players who follow its destructive strategy.

\begin{figure}
\centerline{\epsfig{file=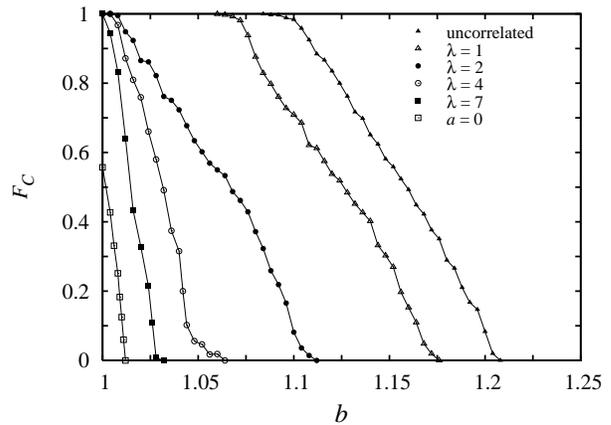,width=8.5cm}}
\caption{\label{fig:fig5}$F_C$ as a function of $b$ for different values of $\lambda$. The $a = 0$ curve is depicted for reference.}
\end{figure}

The above-described feedback works only if the high-ranking defectors can be linked solely by their follower neighborhoods; namely there must not be a strong connection with other high-ranking players. In the opposite case, the heterogeneous strategy distribution supports the survival of defectors. To support our explanation, we study the impact of spatially correlated social diversity on the evolution of cooperation. According to our argument, the transition from uncorrelated to spatially correlated diversity should be marked with the deterioration of cooperation. We thus introduce spatially correlated social diversity as described in Section II, and study the effect of different $\lambda$ on the evolution of cooperation. Figure~\ref{fig:fig5} features the results. In agreement with our conjecture, the facilitation of cooperation deteriorates fast as $\lambda$ increases. Indeed, a near linear decrease of the critical $b$ marking the extinction of cooperators can be established, and moreover, by large $\lambda$ the cooperators can outperform defectors only for slightly higher values of $b$ than in the absence of social diversity ($a = 0$). This result supports the validity of the reported cooperation-facilitating mechanism, and hopefully paves the way for additional studies incorporating social diversity into the theoretical framework of evolutionary game theory.

\section{Summary}
In this paper, we show that social diversity is an efficient promoter of cooperation in the spatial prisoner's dilemma game. The facilitative effect increases with the dispersion of social diversity and deteriorates with the increase of its spatial correlation. Accordingly, spatially uncorrelated power-law distributed social diversity provides the biggest boost to the cooperative strategy. We argue that the facilitative effect is conceptually similar to the one reported previously for inhomogeneous teaching activities, and also has the same root as the mechanism applicable by scale-free networks, namely: akin to hubs or players with enhanced teaching activity, in this model, high-ranking players can form robust cooperative clusters with low-ranking obedient neighbors. The presented results suggest that the distribution of wealth plays a crucial role by the evolution of cooperation amongst egoistic individuals, and it seems reasonable to investigate further whether a co-evolution of both might yield new insights and foster the understanding of the formation of complex societies.

\begin{acknowledgments}
This work was supported by the Slovenian Research Agency (Z1-9629) (M. P.) and the Hungarian National Research Fund (T-47003) (A. S.). Discussions with Gy\"orgy Szab\'o are gratefully acknowledged. We also thank Girish Nathan for useful comments.
\end{acknowledgments}

\end{document}